\documentclass[12pt,titlepage,natbib209,twocolumn]{prop}

\usepackage[letterpaper,margin=1in]{geometry}
\usepackage{html,hyperref}
\usepackage[pdftex]{graphicx}
\usepackage[z-times,expert,alpha]{tmsfnts}
\usepackage{amssymb}

\begin{document}

\title{Towards the 2020 vision of the baryon content\break
 of galaxy groups and clusters}
\author{Andrey Kravtsov\thanks{Dept. of Astronomy \& Astrophysics, Kavli Institute for Cosmological Physics, The University of
Chicago, Chicago, IL 60637, USA; contact author:{\tt andrey@oddjob.uchicago.edu}}, 
Anthony Gonzalez\thanks{Dept. of Astronomy, University of Florida, Gainesville, FL
32611-2055}, Alexey Vikhlinin\thanks{Harvard-Smithsonian Center for Astrophysics, Cambridge, MA 02138; Space Research Institute, Moscow, Russia}, Dan Marrone$^{1,}$\thanks{National Radio Astronomy Observatory, Charlottesville, VA 22903},\\ Ann Zabludoff\thanks{Steward Observatory, University of Arizona, 933 N.Cherry Ave., Tucson, AZ 85721}, Daisuke Nagai\thanks{Dept. of Physics, Yale University, New Haven, CT 06520}, Maxim Markevitch$^{3}$, 
Bradford Benson$^{1}$,\\
Sunil Golwala\thanks{California Institute of Technology
Pasadena, CA  91125}, Steven Myers$^{4}$, Michael Gladders$^{1}$, Douglas Rudd\thanks{Institute for Advanced Study, 1 Einstein Drive, Princeton, NJ 08540},\\ 
August Evrard\thanks{Dept. of Physics, University of Michigan, MI 48109}, Charlie Conroy\thanks{Dept. of Astrophysical Sciences, Princeton University, Princeton, NJ, 08544},
Steven Allen\thanks{Dept. of Physics, Stanford University, Stanford, CA 94305-4060}}

\date{{\it White Paper submitted to the ``Galaxies across Cosmic Time (GCT)''\\ Science Frontier Panel of the Astronomy \& Astrophysics\\ Decadal Survey Committee}\\[1ex]
%%%%%%%%%%%%%%%%%%%%%%%%%%%%%%%%%%%%%%%%%%%%
%% TITLE FIGURE
%%%%%%%%%%%%%%%%%%%%%%%%%%%%%%%%%%%%%%%%%%%%
%\begin{minipage}[h]{6.5in}
%  \begin{center}
%  \includegraphics[width=1.5in]{a1689t.png}
%  \includegraphics[width=1.5in]{spt_as1063_150ghz_t.png}
%  \includegraphics[width=1.5in]{CL6csf1_Stars_t.jpg}
%  \includegraphics[width=1.5in]{cl6_coldfront_t2.png}
%%\\[2ex]
%%  {\scriptsize \it Examples of...}
%  \end{center}
%\end{minipage}
}

\maketitle

%--------------------
\section{The baryon budget of clusters: importance and scientific context}
\label{sec:intro}
%---------------------
 Groups and clusters of galaxies occupy a special position in the
hierarchy of large-scale cosmic structures because they are the
largest and the most massive (from $\approx 10^{13}\,\rm M_{\odot}$ to
over $10^{15}\,\rm M_{\odot}$) objects in the universe that have had
time to undergo gravitational collapse. The large masses of clusters
imply that their contents have been accreted from regions of $\approx
8-40$~comoving Mpc in size and should thus be representative of the
mean matter content of the universe \cite{white_etal93}. Thus, in
contrast to galaxies, clusters are expected to retain the cosmic
fraction of baryons within radii accessible to current and next
generation observations in sub-mm, optical/NIR and X-ray wavelengths,
even if their diffuse gas was significantly heated by supernovae (SNe)
and active galactic nuclei (AGN).

The gaseous atmospheres of clusters (the intracluster medium or ICM)
therefore contain a wealth of information about galaxy formation
processes, including the efficiency with which intergalactic gas is
converted into stars and the chemical and thermodynamical effects on
the evolution of galaxies and the surrounding intergalactic medium
(IGM) driven by ensuing feedback mechanisms. Suppose, for example,
that AGN feedback significantly heats the ICM in the central regions
and the gas expands.  The density of the gas in the center would be
lower, but we should still observe the expanded gas in the outer
regions of the cluster and should be able to observe the signature of
heating in its entropy
\cite{bialek_etal01,bower_etal08,bhattacharya_etal08}.  Likewise, if a
certain fraction of ICM gas has cooled in the past, we should be able
to observe these baryons either in the form of cold gas or in the form
of stars \cite{kravtsov_etal05}.  Some fraction of gas processed by
galaxies is returned to the intracluster medium via winds and ram
pressure stripping, as is evidenced by the high abundance of heavy
elements in the intracluster plasma. The distribution of metals as a
function of radius provides a record of the baryons recycled through
galaxies and the exchange of material between galaxies and surrounding
gas. Multi-wavelength, holistic studies of baryon contents of groups
and clusters {\it as a function of radius} thus shed light on many key
processes accompanying galaxy formation, one of the central unsolved
problems of modern astrophysics. Most of these processes are difficult
to observe or constrain by other means.

Knowledge of the baryon content of clusters is also a key ingredient
in the use of clusters as cosmological probes. First, it is the
observed stellar and hot gas components of clusters that are most
often used as proxies for their total mass, which can be connected to
theoretical predictions and used to constrain cosmology via evolution
of the cluster abundance and spatial distribution
\cite{vikhlinin_etal09b}.  Second, measuring the fraction of the total
mass in baryons within clusters is one of the most powerful methods of
measuring the mean matter density of the universe
\cite{white_etal93,evrard97} and is important in the interpretation of
the cluster signal contribution to the small-scale CMB anisotropies in
terms of the power spectrum normalization
\cite{white_etal02,komatsu_seljak02,sharp_etal09,sievers_etal09}.
Third, since the mass fraction of hot gas measured from observations
depends on the luminosity distance at the cluster redshift, tracing
evolution of this fraction as a function of redshift is a powerful
independent geometrical constraint on the dark energy content of the
universe
\cite{sasaki96,pen97,allen_etal02,ettori_etal03,laroque_etal06,allen_etal04,allen_etal08}. All
of the above measurements require good understanding of the evolution
of the stellar and hot gas fractions in clusters.

  Study of the baryon budget of groups and clusters across cosmic time
is thus a truly important scientific theme with a wealth of
connections to other areas of astronomical research, such as
cosmology, galaxy formation and evolution, AGN physics, etc., and is
therefore of tremendous scientific interest. During the past two
decades, the advent of new observational facilities and theoretical models
have allowed us to uncover some very interesting puzzles in this area
(see \S~\ref{sec:current}) and to formulate specific questions, which
we are poised to address in the next decade:\\
\begin{list}{$\rhd$}{
  \setlength{\topsep}{0ex}\setlength{\itemsep}{0ex plus0.2ex}
  \setlength{\parsep}{0.5ex plus0.2ex minus0.1ex}}
\item {\it Does the baryon fraction of groups and clusters reflect that of the universe as a whole at radii and epochs probed by observations?} 
\item {\it What are the radial distributions of different mass components 
of clusters (stars and cold gas, diffuse hot gas, and dark matter)?}
\item {\it How does the observed baryon budget in clusters
and groups constrain models of galaxy formation and
feedback?}\\
\end{list}

The radial distributions of all main mass components (stars, cold,
warm, and hot gas and total mass) at different redshifts out to the
virial radius {\it in the same clusters} are the main required
observational measurements. During the next decade sensitive
multi-wavelength observations, capable of accurately probing all the
major matter components of clusters, should allow us to tackle or make
substantial progress on all of the outlined questions.  Increasingly
sophisticated simulations of clusters will explore a range of
plausible scenarios and assumptions about the physics of galaxy
formation and ICM and their effect on resulting distribution of
baryons and total mass.  Multi-wavelength, comparative studies of real
and simulated cluster samples would allow us to use clusters as
veritable astrophysical laboratories for studying galaxy formation and
testing our theoretical models of structure formation and underlying
assumptions about fundamental physics governing the
universe.\footnote{The Bullet cluster, the system in which hot gas in
one of the merging clusters is clearly offset from center of total
mass and galaxies, is a fine example of a system where
multi-wavelength mapping of different components of cluster region
provides powerful information about underlying physics of the system
and properties of dark matter \cite{clowe_etal06}.}  At the same time, reliable and
detailed knowledge of the baryon content of clusters would provide the
necessary basis for their use as precision cosmological probes.

%---------------------------------
\section{Current status}
\label{sec:current}
%---------------------------------

\subsection{Observations.} 
Over the past decade, dramatic progress has been made in observations
of all the main components of clusters. Deep high-resolution wide-field imaging 
has dramatically improved the accuracy of strong and weak lensing measurements of
cluster mass profiles and the number of systems with lensing data
has been increasing dramatically \cite{mahdavi_etal08,zhang_etal08}. Such measurements provide a powerful 
independent method of measuring the distribution of total mass in clusters. 
Statistical measurements of the shear profiles around large numbers of clusters
in the SDSS survey have determined the average mass profiles around
clusters of different richness with exquisite precision out to $\sim 10$~Mpc
 \cite{sheldon_etal07}. 

Deep imaging in multiple optical bands, in particular in the near
infrared (NIR), has permitted a systematic census of the stellar
component in clusters \cite{lin_etal03,gonzalez_etal07},
 including the
low surface brightness intracluster light (ICL). At the same time, measurements
of stellar kinematics with the panoramic integral field spectrograph
SAURON \cite{cappellari_etal06} 
have provided the first direct constraints on the
mass-to-light ratios of stellar populations of some nearby cluster
galaxies.

New, high-resolution, sensitive X-ray observations by {\sl Chandra\/}
and {\sl XMM}-Newton have revolutionized our knowledge about
properties of the hot ICM gas by reliably mapping gas
density, temperature, and metallicity out to a significant fraction of
virial radius \cite{vikhlinin_etal06,pratt_etal07}.
 These observations show that the ICM is not isothermal
and that the widely used $\beta$-model does not describe the radial
distribution of the hot gas density in the outer regions of
 clusters. Qualitatively better data about the thermal state and cooling of the
gas, especially in cluster cores, has revealed both new insights and
puzzles, such as lack of very cold gas in the centers \cite{peterson_fabian06} and high gas entropy
in the outskirts \cite{ponman_etal03}.  Deep, targeted X-ray observations of
carefully selected relaxed clusters have determined the total
mass profiles of groups and clusters under the assumption of hydrostatic
equilibrium and put constraints on the halo
concentrations and cuspy inner density profiles predicted by Cold Dark
Matter (CDM) structure formation simulations \cite{pointecouteau_etal05,vikhlinin_etal06,buote_etal07}. 

Measurements of the Sunyaev-Zeldovich effect (SZE) with
interferometric techniques have produced independent constraints on
the thermal state of the gas in dozens of clusters
\cite{laroque_etal06,bonamente_etal08}.  At the same time,
technological breakthroughs in building large bolometric detector
arrays have improved the mapping speed of the SZE receivers by an
order of magnitude \cite{ruhl_etal04,fowler_etal07}. Such instruments
now provide unique constraints on the ICM gas distribution in the
outermost regions of clusters, to the virial radius and beyond.

\begin{figure}
%\vspace{-2.5cm}                                                                                                                                                                  
%\hspace{0.75cm}\centerline{{\includegraphics[width=0.35\linewidth]{gzz07_f5.png}}                                                                                                
%{\hspace{-0.2cm}\includegraphics[width=0.4\linewidth]{spt_as1063_150ghz_radial_profile.png}}                                                                                     
%{\hspace{-0.2cm}\includegraphics[width=0.45\linewidth]{chandra_rhoprof.pdf}}}                                                                                                    
\centerline{\vspace{-0.35cm}\includegraphics[width=0.9\linewidth]{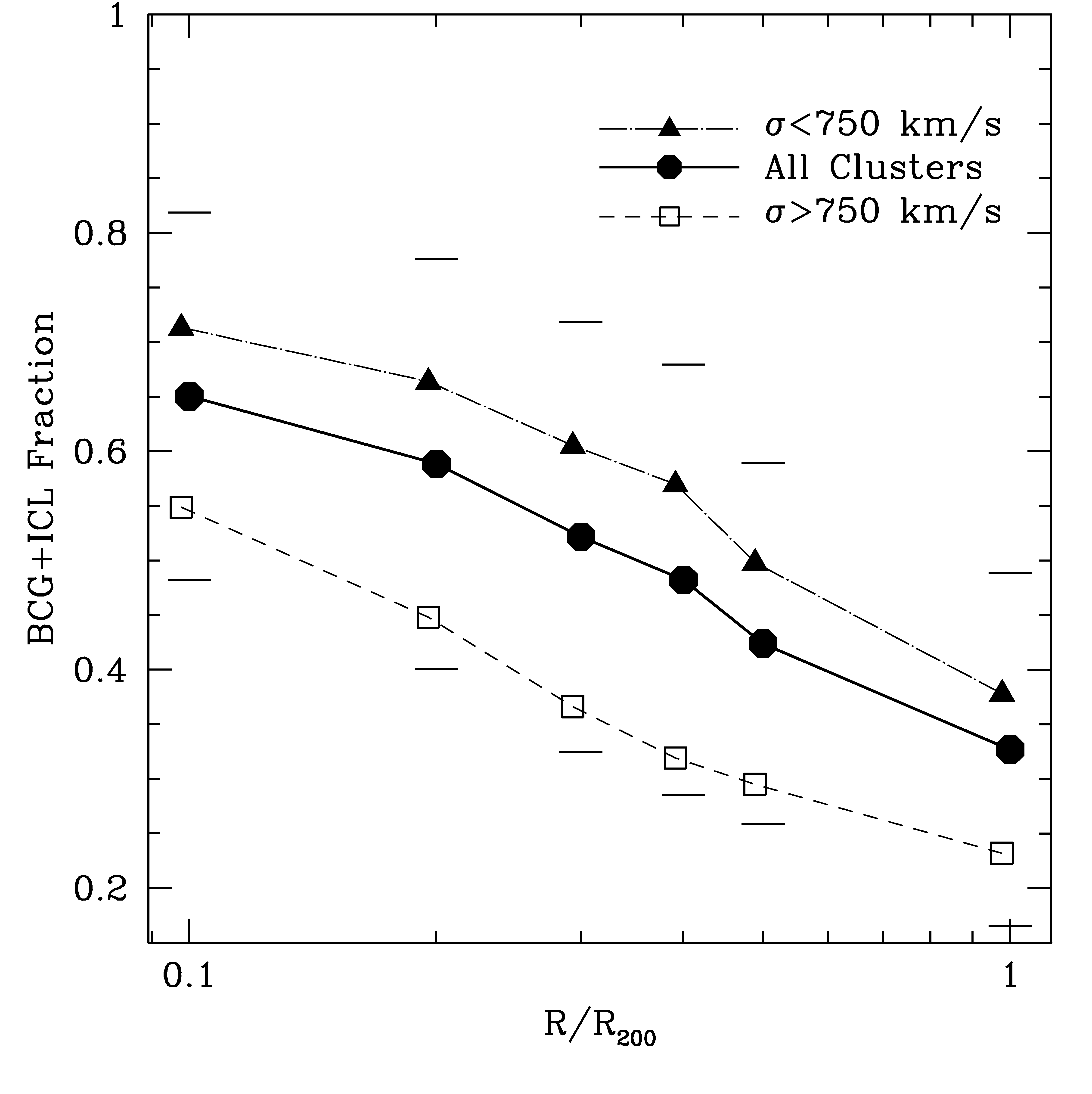}}
\centerline{\vspace{-5.25cm}\hspace{-0.2cm}\includegraphics[width=0.98\linewidth]{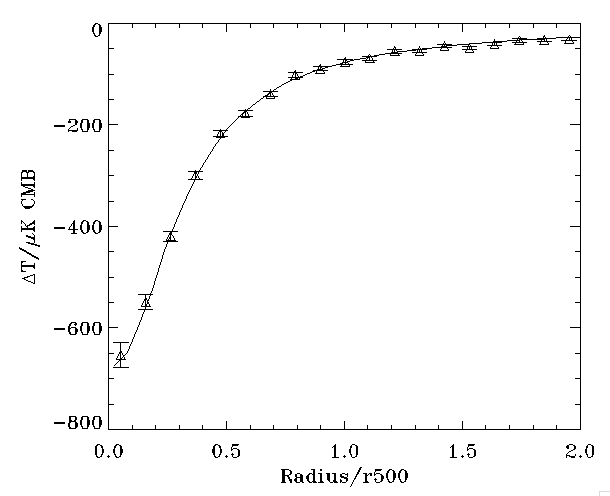}}
\centerline{\vspace{-0.5cm}\hspace{2.5cm}\includegraphics[width=1.2\linewidth]{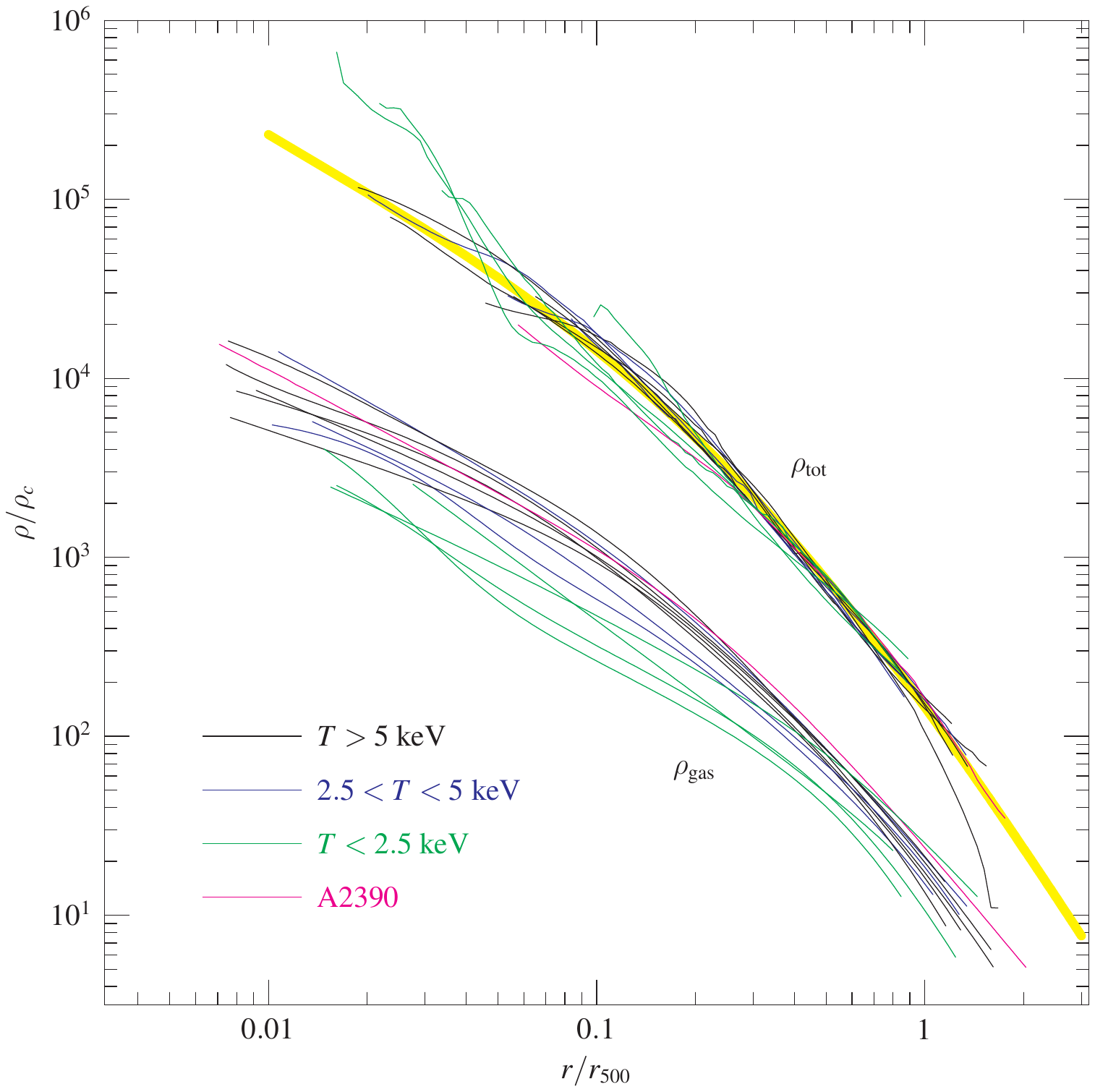}}
%\vspace{-6cm}                                                                                                                                                                    
\caption{Distribution of mass fraction of stars in the brightest
cluster galaxies and ICL in clusters of different mass
(top,\protect\cite{gonzalez_etal07}), accurate SZE profile measurement
for cluster AS1063 out to the virial radius, $\approx 2r_{500}$
(middle, SPT team, in prep.), and hot gas and total mass profiles
derived from deep {\sl Chandra} data for a sample of relaxed clusters
(bottom, \protect\cite{vikhlinin_etal06}).
}
\label{fig:obsex}
\end{figure}

Deep 21 cm imaging of the Virgo cluster has constrained the amount of
cold HI gas \cite{chung_etal08}, while optical and molecular
observations have revealed the presence of some cold gas in cluster
centers \cite{edge01,salome_combes03,odea_etal08,wilman_etal09}. These measurements have started to
elucidate the way in which the interstellar medium of galaxies gets
stripped by the hot ICM as they enter clusters and the balance of
heating and cooling in cluster cores.  UV spectroscopy of background
quasars has given first constraints on the amount and properties of
warm gas ($T\sim {\rm few}\times 10^5$~K).
 
Figure~\ref{fig:obsex} shows three representative examples out of many exciting
developments outlined above: measurement of the contribution of the central galaxy and ICL stars to 
the total stellar budget of clusters out to the virial radius, recent high 
signal-to-noise measurements of the SZE signal out to the virial radius, 
and measurements of ICM gas density profiles and total mass profiles from deep
targeted {\sl Chandra} X-ray observations of relaxed clusters. The figure demonstrates
that current observations are becoming capable of  mapping profiles of different cluster 
components to a significant fraction of the virial radius.  

%------------------
\subsection{Theory.}
%------------------
During the last two decades self-consistent cosmological simulations
of cluster formation following the dynamical evolution of
collisionless dark matter and diffuse baryons in the non-radiative
regime and including various processes accompanying galaxy formation
have improved in mass resolution and dynamic range by about 4 and 2
orders of magnitude, respectively. Despite the fact that many aspects
of galaxy formation remain poorly understood, notable successes of the
simulations include correct {\it predictions} of the gas density and
temperature profiles outside of cluster cores
\cite{evrard90,frenk_etal99,nagai_etal07b}, and the radial distribution
of cluster galaxies \cite{nagai_kravtsov05}. Prediction of these
quantities in an ab initio model of $\Lambda$CDM structure formation
is convincing evidence that modeling of cluster formation is on a firm
theoretical footing.  Cosmological simulations of clusters have
demonstrated that the baryon fraction outside cluster cores is
expected to be within $\approx 10\%$ of the universal baryon fraction
$\Omega_{\rm b}$.  \cite{frenk_etal99,kravtsov_etal05,crain_etal07}.
However, stellar and hot gas fraction profiles do depend sensitively
on the details of the physical processes included in simulations.
\cite{bialek_etal01,kravtsov_etal05,ettori_etal06,bower_etal08}.

\begin{figure}[t]
%\vspace{-2.5cm}
%\hspace{1.2cm}\centerline{\includegraphics[width=1.1\linewidth]{fbgm_wp.pdf}}
%\vspace{-6cm}
\vspace{-2.25cm}
\hspace{1.0cm}\centerline{\includegraphics[width=1.5\linewidth]{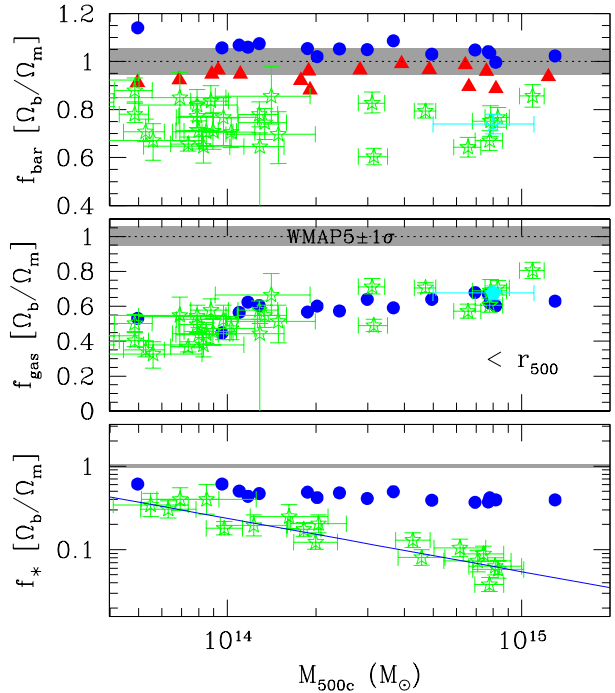}}
\vspace{-4.3cm}
\caption{The stellar (bottom), gas (middle) and total baryon (top) mass
fractions in units of the mean baryon fraction ($\Omega_{\rm b}/\Omega_{\rm m}$) 
in cosmological cluster simulations with (blue) and without (red) galaxy 
formation (points with no error bars) and in observed clusters \protect\cite{vikhlinin_etal06,gonzalez_etal07,zhang_etal06,sun_etal08} 
(points with error bars). 
within the radius $r_{500}$ enclosing overdensity of $500$ times the critical
density. The figure shows fairly good agreement of observed and predicted gas fractions, 
but large discrepancies in the stellar fractions. The total baryon
fractions of most of the observed clusters are significantly below the theoretical
expectation.  }
\label{fig:fbar}
\end{figure}

Figure~\ref{fig:fbar} illustrates the current observational
measurements of stellar and hot gas mass fractions, and their sum --
the total baryon fraction, within $r_{500}$ (the radius enclosing an
overdensity of $500\rho_c$, $\rho_c\equiv 3H(z)^2/8\pi G$, $r_{500}\approx 0.5r_{\rm
vir}$), and compares these measurements to high-resolution
simulations of clusters with dissipative physics of galaxy formation
\cite{nagai_etal07b}.

Stellar mass fractions in observed groups and clusters (bottom panel) steadily decrease
with increasing cluster mass. Clusters from cosmological simulations shown in the figure
exhibit at best a very weak trend. The actual values of $f_{\ast}$ for groups ($M<10^{14}\,\rm M_{\odot}$)
are in reasonably good agreement. For massive clusters, however, the $f_{\ast}$
in simulated clusters is a factor of $\sim 2-5$ larger than observed. 
This is an illustration of the well-known ``overcooling problem.''\footnote{This problem
should be distinguished from the ``cooling flow'' problem \cite{peterson_fabian06}. The latter refers
to the lack of cold gas in cluster centers, while the overcooling 
problem refers to  a too high fraction of cooled, condensed gas
produced in simulations {\it over the entire formation history} of clusters,
including all of its progenitors.} 

Surprisingly, despite the glaring discrepancy in $f_{\ast}$, the comparison of
hot gas fractions in the middle panel shows rather good agreement
between simulations and data, both in the average trend and in the
actual values. 
% Note that the gas fraction in the simulation with no cooling at
%$z<2$  is considerably higher than the mean observed
%fraction at the corresponding mass.  
%This means that suppressing the
%stellar fraction without heating the gas, so as to push a significant
%fraction of it outside $r_{500}$, is not a viable solution to the
%overcooling problem.
The top panel shows comparisons of the total baryon fractions
in simulations performed both in non-radiative regime (red
triangles) and with galaxy formation\footnote{The $f_{\rm bar}$ values in
these two types of simulations are close to the universal but are somewhat different. Physical
processes behind these differences are understood, but their explanation is beyond the scope of this paper.} (blue circles) to observations. 
The total baryon fractions of both observed and simulated clusters do not show a detectable trend with mass. 
For observed systems this, by itself, is a striking fact as 
gas in the shallower potential wells of groups is expected to be more susceptible to AGN feedback.
Another noteworthy fact is that observed baryon fractions  are lower 
than the universal value and the values predicted by simulations for this radius
\cite{ettori03,afshordi_etal07,mccarthy_etal07b}. 
Although there is a
scatter in values of $f_{\rm bar}$ from cluster to cluster, a
large fraction of them is significantly below the universal
value measured by the {\sl WMAP\/} satellite.

There are several different potential explanations for the low
observed baryon fractions, although none of them have been
convincingly proven yet. One key theoretical issue is
systematic errors in stellar masses derived from stellar population synthesis models \cite{maraston_etal06,conroy_etal09}, 
but other unknown systematics in estimates of stellar  and gas masses are possible. The X-ray estimates
of the latter, for example, depend on the Hubble constant as $f_{\rm gas}\propto h^{-1.5}$ so a value
of $h=0.65$ ($2\sigma$ lower than $h=0.72$ -- the best fit value to the WMAP 5 yr data) would explain
a significant fraction of the difference.

The discussed comparisons of expectations and observations
 show that our current ``vision'' and understanding of the baryon budget within 
virial radius are not yet as clear as we would like them to be. 
The possibility remains that we may be missing key details of the picture. However, the situation can 
be dramatically improved in the next decade with the advent of a new generation
of surveys and instruments at different wavelengths. 

%---------------------------------
\section{Studying baryon budget in the next decade}
%---------------------------------

%---------------------------------
\subsection{Observations.}
%---------------------------------
The next generation of observations to constrain baryon budget of
groups and clusters should cover a representative range of redshifts
($0<z<2$). This would allow to study formation of these systems as it occurs (and where
sensitivity to dark energy is greatest in cosmological tests). Additionally, observations should probe
a wide range of masses, as smaller systems merge and are incorporated
into larger ones, so their properties are key to understanding of the
baryon content of the larger systems. Lastly, it is  very important to
trace mass components as far in radius as possible, as cluster
outskirts may contain key information both about accretion of matter
onto clusters and about the prior thermal history of baryons.  The
observational program can address the scientific questions formulated in
\S~\ref{sec:intro} by a combination of statistical studies of clusters
in large wide-area surveys to study mean trends and deep targeted
observations of individual systems to study variations as a
function of specific physical properties.

{\it Stellar content.} To obtain robust stellar mass measurements for
cluster galaxies and ICL, deep, multi-band photometry is
needed, with filters spanning the peak in the stellar component of the
SED. This implies a combination of optical and infrared imaging for
the redshifts $0<z<2$. Such imaging can be carried out efficiently
using a combination of a large wide area optical ground-based or space
imaging survey, and NIR imaging with JWST from space. 
Large area multi-band optical surveys
could constrain ICL in low-$z$ clusters for which X-ray and SZ data are available by
stacking clusters within a given range of
total mass. Independent constraints on the mass-to-light ratios of a
large number of cluster galaxies with IFU spectroscopy on large
telescopes would usefully extend upon the work of SAURON project in
the recent years. Development of better SPS models and stellar mass
estimators and good understanding of the associated systematic error is
a must \cite{conroy_etal09}. 
Spectroscopic measurements of velocities in cluster fields
in conjunction with theoretically-motivated mock catalogs could
constrain the 3d distribution of cluster
galaxies. Efficient execution of such a spectroscopic measurement
program requires a highly multiplexed spectrograph capable of dense
spatial sampling (e.g.  an instrument like GISMO+IMACS at Magellan) or
a wide-area IFU, on an $8-10$~m or larger telescope. 

{\it Hot ICM gas.} Targeted deep observations of individual groups and
clusters by {\sl Chandra,\/} {\sl XMM}-Newton, and {\sl Suzaku\/} will
continue to provide measurements of gas density and temperature
profiles out to $r_{500}-r_{\rm vir}$. The number of possible targets
will increase dramatically after completion of the currently ongoing
X-ray and SZ surveys and those planned for the near future. It would
be extremely helpful if by the end of the next decade a more
sensitive, low background, high-resolution ($<10^{''}$) X-ray
telescope capable of mapping the gas distribution to larger radii (to
$r_{\rm vir}$ and perhaps beyond) at $0<z<2$, such as {\sl IXO} 
(see white paper by Vikhlinin et al.), would become available.  Sensitive X-ray observations
would help to trace evolution of hot gas fraction very accurately over
the main epochs of cluster formation and for a wide range of masses,
which would provide a larger lever arm for comparisons with
models. They would also provide robust estimate of total cluster mass using
established robust, low-scatter observable mass proxies \cite{kravtsov_etal06}.

SZE observations of hot ICM will play an increasingly important role
in the next decade (see white paper by S. Golwala et al.), as a
combination of new sensitive interferometers and single-dish telescopes
with large, multi-band $\sim mm$ focal plane detector arrays should be
coming online. The unprecedented combination of high angular
resolution ($\sim$arcsecond) and spatial dynamic range of the new
planned interferometers will allow detailed imaging of the ICM thermal
pressure within inner regions of clusters ($\theta<1^{\prime}$) with
resolution well matched to that of the {\sl Chandra\/} X-ray
observations. Such high-resolution observations will also be able to
estimate the contribution of background sources to the SZE signal,
thereby eliminating one of the main sources of systematic
uncertainty. Continuing development and deployment of SZ receivers with large bolometric detector arrays
($>10^4$ detectors) on telescopes with wide fields is critical to map
out thermal pressure of the ICM to the virial radius and beyond 
 for representative samples of clusters. 

The combination of X-ray and new generation of SZE observations in the
next decade would be particularly powerful in mapping the distribution
and thermal properties of the hot ICM out to large radii and in
cross-checking for systematics. Such observations would decisively
constrain or exclude various scenarios of galaxy formation and
feedback in clusters, as well as plasma effects, such as helium
sedimentation
\cite{chuzhoy_loeb04,markevitch07,ameglio_etal07,peng_nagai09}.

{\it Mapping the cold gas.} Mass fractions of cold gas are small
compared to stars and hot gas, but their measurement is important for
understanding just how multiphase the ICM really is. Such measurements
will also provide information on the effectiveness of feedback from
galaxies and AGN by constraining how much of the central gas actually
cools. During the next decade ALMA will revolutionize measurements of
molecular gas in clusters over a wide range of redshifts, while Square
Kilometer Array pathfinder missions can be used to map diffuse HI gas
in distant clusters.

{\it Mapping the total mass distribution.} There are now several
well-established techniques for deriving the distribution of total
mass in clusters.  Accurate gas distribution and temperature profiles
obtained with existing and future X-ray observatories can provide
estimates of the total mass profile for hundreds of relaxed clusters
at different redshifts. At the same time, the availability of
high-resolution SZ observations would allow application of such
techniques to more distant clusters and provide useful cross-checks
for lower $z$ clusters \cite{ameglio_etal07}, for which both X-ray and
SZ data will be available.  Additional checks will come from
measurements of strong lensing masses in cluster cores \cite{lemze_etal08,oguri_etal09}, as statistical
samples of arcs emerge from the upcoming wide area photometric surveys
with follow up from space.

Wide-field deep imaging from large ground-based telescopes, but
especially by imaging cameras on space missions, will allow strong and
weak lensing measurements of total mass profiles (with potential great
synergy between the two). Such measurements for large samples of
clusters will provide independent mass measurements and will allow
statistical calibration of masses derived from hydrostatic equilibrium
analyses of X-ray and SZ data.

%---------------------------------
\subsection{Theoretical modeling.}
%--------------------------------                           
{\ }The concerted, multi-wavelength observational campaign outlined
above will need to be accompanied by improvements in theoretical
modeling of cluster baryons. Specific, robust theoretical predictions
of the properties of cluster galaxies, profiles of ICL, and gas
density distribution and thermal structure of the hot ICM are
required.  Simulations are not yet at the point where they can
reliably predict properties of cluster galaxies or faithfully model
various aspects of galaxy formation and ICM evolution. The focus will
therefore be on exploring a wide range of plausible scenarios and
different physical processes: efficiency of star formation and its
dependence on environment, efficiency of stellar and AGN feedback,
generation and evolution of non-thermal components in the form of
cosmic rays and magnetic fields (see white papers by Myers et al. and
Rudnick et al.), etc.  Progress is particularly needed in
understanding the role of collisionless effects and deviations from
equilibrium in plasma. This will be essential for reliable
interpretation of hot gas fraction measurements, especially in the
cluster outskirts.

At the same time, advances in available computing power should allow
for larger simulations which will better resolve small-scale processes
during galaxy formation. Note that cosmological simulations of 
clusters are particularly difficult in this respect, as scales of kpc (or
smaller) need to be resolved to model the ISM and star formation in galaxies
reasonably, while modeling volumes in excess of 100 Mpc. 
To simply resolve the ICM within a virial radius of a massive cluster uniformly to 1 kpc
scale requires $\sim 10^{11}$ resolution elements, well beyond
capability of the current simulations. Modern methods overcome this daunting
requirement by employing either Lagrangian or adaptive mesh refinement techniques,
which concentrate resolution elements in small fractions of the volume in
cluster center and around densest regions of cluster galaxies. Resolving
small-scale turbulence, shocks, cold fronts, and other potentially important
phenomena over large volume will challenge the capabilities of these techniques. The wide
range of included physical processes also significantly increases requirements
for memory, storage, and bandwidth capabilities. This means
that substantial tera- and peta-scale computing facilities will be required
for successful work in this direction and argues for continuing investment 
in such facilities. Investment into funding computer science research is also
needed, as it would help ensure that simulation codes can effectively use
available computing platforms.

\bibliography{barwp}

\end{document}